# Discovery of an Antiferromagnetic Topological Nodal-line Kondo Semimetal

D. F. Liu[1,2*†], Y. F. Xu[3*†], H. Y. Hu[4*], J. Y. Liu[5,6*], T. P. Ying[7*], Y. Y. Lv[8*], Y. Jiang[4*], C. Chen[9], Y. H. Yang[5], D. Pei[5], D. Prabhakaran[5], M. H. Gao[8], J. J. Wang[7], Q. H. Zhang[7], F. Q. Meng[10], B. Thiagarajan[11], C. Polley[11], M. Hashimoto[12], D. H. Lu[12], N. B. M. Schröter[13], V. N. Strocov[13], A. Louat[6], C. Cacho[6], D. Biswas[6], T.-L. Lee[6], P. Steadman[6], P. Bencok[6], Y. B. Chen[8], L. Gu[10], T. Hesjedal[5], G. van der Laan[6], H. Hosono[14], L. X. Yang[15], Z. K. Liu[9], H. Q. Yuan[3], B. A. Bernevig[4,16†], Y. L. Chen[5,9†]

[1]*School of Physics and Astronomy, Beijing Normal University, Beijing, 100875, China*
[2]*Key Laboratory of Multiscale Spin Physics, Ministry of Education, Beijing, 100875, China*
[3]*Center for Correlated Matter and School of Physics, Zhejiang University, Hangzhou, 310058, China*
[4]*Donostia International Physics Center (DIPC), Paseo Manuel de Lardizábal, 20018, San Sebastián, Spain*
[5]*Department of Physics, Clarendon Laboratory, University of Oxford, Oxford OX1 3PU, UK*
[6]*Diamond Light Source, Harwell Science and Innovation Campus, Didcot OX11 0DE, UK*
[7]*Institute of Physics, Chinese Academy of Sciences, Beijing, 100190, China*
[8]*Department of Materials Science and Engineering, Nanjing University, Nanjing, 210093, China*
[9]*School of Physical Science and Technology, ShanghaiTech University, Shanghai, 201210, China*
[10]*Departement of Materials Science and Engineering, Tsinghua University, Beijing, 100084, China*
[11]*MAX IV Laboratory, Lund University, Lund, 22484, Sweden*
[12]*Stanford Synchrotron Radiation Light Source, SLAC National Accelerator Laboratory, Menlo Park, California, 94025, USA*
[13]*Swiss Light Source, Paul Scherrer Institute, Villigen,5332, Switzerland*
[14]*Materials Research Center for Element Strategy, Tokyo Institute of Technology, Yokohama, 226-8503, Japan*
[15]*Department of Physics, Tsinghua University, Beijing, 100084, China*
[16]*Department of Physics, Princeton University, Princeton, NJ, 08544, USA*

\* These authors contributed equally to this work
†Email: defa.liu@bnu.edu.cn; y.xu@zju.edu.cn; bernevig@princeton.edu; yulin.chen@physics.ox.ac.uk

**The symbiosis of strong interactions, flat bands, topology and symmetry has led to the discovery of exotic phases of matter, including fractional Chern insulators, correlated moiré topological superconductors, and Dirac and Weyl semimetals[1-14]. Correlated metals, such as those present in Kondo lattices, rely on the screening of local moments by a sea of non-magnetic conduction electrons[15,16]. Here, we report on a unique topological Kondo lattice compound, $CeCo_2P_2$, where the Kondo effect – whose existence under the magnetic Co phase is protected by $P·T$ symmetry – coexists with antiferromagnetic order emerging from the flat bands associated with the Co atoms. Remarkably, this is the only known Kondo lattice compound where magnetic order occurs in non-heavy electrons, and puzzlingly, at a temperature significantly higher than that of the Kondo effect.**

**Furthermore, at low temperatures, the emergence of the Kondo effect, in conjunction with a glide-mirror-*z* symmetry, results in a nodal line protected by bulk topology near the Fermi energy. These unusual properties, arising from the interplay between itinerant and correlated electrons from different constituent elements, lead to novel quantum phases beyond the celebrated topological Kondo insulators[17-24] and Weyl Kondo semimetals[4-14]. $CeCo_2P_2$ thus provides an ideal platform for investigating narrow bands, topology, magnetism, and the Kondo effect in strongly correlated electron systems.**

Over the past decades, the successful application of topology in band theory has led to the discovery of a wide variety of topological quantum materials in weakly interacting systems, such as topological insulators (TIs)[25-31], topological superconductors (TSC)[30-33], Dirac semimetals (DSMs)[34-37] and Weyl semimetals (WSMs)[36-42]. These phases host exotic bulk fermions and robust boundary states, resulting in unique physical properties such as dissipationless edge current[43], giant magneto-transport[44] and axion electrodynamics[45]. Moreover, these materials also hold significant potential for applications ranging from low power electronics[46], novel spintronics[47], efficient thermoelectric materials[48] to fault-tolerant quantum computation[49].

Despite significant progress in non-interacting systems, three-dimensional (3D) phases in strongly correlated electron systems that mix interactions and topology have been less explored, as they are challenging to predict theoretically and manipulate experimentally. Interactions are naturally enhanced in flat (narrow) band systems, where the discoveries of two-dimensional (2D) fractional Chern insulators[1] and superconductivity in highly tunable moiré systems[2] have taken condensed matter physics by storm. These advances suggest the presence of exotic new phases in 3D, if flat (narrow) bands, strong interactions, symmetry and topology can be brought together in a controllable way.

Recent advances in heavy-fermion materials have led to the discovery of new topological states in Kondo systems[4-14,17-24]. The Kondo effect, a hallmark of strongly correlated electron phenomena, arises from the scattering of spinful nonmagnetic conduction electrons by local magnetic moments and is typically observed in rare-earth compounds with partially filled *f*-orbitals[15,16]. Kondo systems, and especially the less understood Kondo lattices, can exhibit intriguing properties such as unconventional superconductivity[50,51], non-Fermi liquid behavior[52-55], and most recently topological Kondo insulators (TKIs)[17-24]. Recently proposed Kondo semimetals[4-13] have been indirectly investigated through transport measurements, displaying giant spontaneous Hall effect[10] and colossal anomalous Nernst effect[11]. When the conduction electrons also exhibit narrow band portions at the Fermi level, their enhanced density of states could lead to many-body states, whose interplay with the Kondo effect remains unexplored; and the effects of the system's symmetries on these properties are still poorly understood.

In this work, we demonstrate that the low-carrier-density Kondo lattice compound $CeCo_2P_2$ is unique among all known Kondo lattice compounds to date. It is the only known compound in nature where the Kondo effect emerges deep in an antiferromagnet – a unique case where the Kondo effect coexists with the magnetic ordering of narrow bands of Co conduction electrons, raising the intriguing question of what screens the Ce moments. Theoretically, we show that this phenomenon is due to a remaining $P \cdot T$ symmetry of this magnetic system, which guarantees the presence of "nonlocal" Kramer's doublets that now act as a conduction isospin[56]. Through angle-resolved photoemission spectroscopy (ARPES) and transport measurements, we reveal that the emergence of the Kondo effect, in cooperation with the crystal's glide-mirror-$z$ symmetry, gives rise to a nodal-line Fermi-surface (FS) protected by the bulk topology near the Fermi energy ($E_F$). These unusual properties, arising from the interaction between itinerant and correlated electrons from different constituent elements (Fig. 1a), make $CeCo_2P_2$ an ideal platform for investigating the interplay between topology, magnetism, and Kondo effect in strongly correlated electron systems.

$CeCo_2P_2$ crystallizes in a body-centered tetragonal lattice (space group: *I4/mmm*) with alternating Ce and [P-Co$_2$-P] layers (Fig. 1b). The high quality of the single crystal samples used in this study is confirmed by high-resolution scanning transmission electron microscopy (STEM) imaging (Fig. 1c), where both the Ce and [P-Co$_2$-P] layers are clearly visible (inset of

Fig. 1c). In each [P-Co$_2$-P] layer, the Co atoms form a 2D square lattice that magnetically orders at a Néel temperature of approximately 440 K (Fig. 1d, red curve). Within each layer, the Co magnetic moments are ferromagnetically (FM) ordered; while anti-ferromagnetic (AFM) ordering is formed between adjacent Co layers, as confirmed by neutron scattering studies[57] (illustrated in the inset of Fig. 1d). The in-plane FM and out-of-plane AFM order can be understood through layer-narrow-band ferromagnetism stabilized by quantum geometry, and out-of-plane antiferromagnetic Heisenberg coupling which results from the absence of out-of-plane quantum geometry, respectively[56].

Remarkably, in its AFM state, CeCo$_2$P$_2$ exhibits Kondo-like behavior similar to that of other canonical Kondo systems[58]. As shown in Fig. 1d, upon decreasing the temperature, the resistivity measurements display a typical Kondo upturn [$\rho \sim -\ln(T)$], peaking at the coherence temperature ($T^*$), followed by a rapid decrease and a return to Fermi-liquid behavior ($\rho \sim T^2$) at low temperatures (Extended Data Fig. 1). The Kondo effect arises from the $4f^1$ electrons of the Ce$^{3+}$ atoms, as confirmed by X-ray absorption spectroscopy (XAS, Fig. 1e) and X-ray photoemission spectroscopy (XPS, Fig. 1f), both of which show dominant characteristic peaks of Ce$^{3+}$ (Extended Data Fig. 5).

The emergence of the Kondo effect, in conjunction with glide-mirror-$z$ symmetry, further establishes CeCo$_2$P$_2$ as a nodal line semimetal protected by bulk topology[56,60]. Our calculations show that a band inversion occurs between the Co-3$d$ and Ce-4$f$ bands at the Λ point near $E_F$, leading to the formation of a nodal line in the bulk Brillouin zone (BZ) at $k_z = \pm 0.33$ Å$^{-1}$ (green rings in Fig. 1g and 1h; Extended Data Fig. 4 and Refs. 56, 60). Hall measurements indicate a carrier concentration of ~ 10$^{21}$ cm$^{-3}$, confirming the low-carrier-density semi-metallic properties, consistent with the nodal line semimetal state (Extended Data Fig. 3). These findings underscore the importance of narrow bands, lattice symmetry, quantum geometry and topology in understanding correlation physics (Fig. 1a), as we will further discuss below.

To illustrate the electronic structure and its temperature evolution during the Kondo transition, we carried out systematic ARPES measurements, including the resonance ARPES near the Ce 4$d$ → 4$f$ absorption edge (~ 121 eV) to enhance the $f$-electron spectral intensity (Fig.

2a). Using this on-resonance photon energy, the *f*-states from Ce are clearly visible in the vicinity of $E_F$ [Fig. 2b (ii)], corresponding to the $4f^1_{5/2}$ and $4f^1_{7/2}$ levels[61,62] [Fig. 2b (iii)], respectively. This allowed us to perform detailed temperature-dependent measurements and analysis, as presented in Fig. 2c and 2d.

As shown in Fig. 2c, upon lowering the temperature, the $4f^1_{5/2}$ band of Ce gradually emerges at $E_F$, characterized by the growth of sharp quasi-particle peaks in the energy distribution curves (EDCs). The inset of Fig. 2c displays two dispersion plots from ARPES measurements taken well above and below the coherence temperature ($T^* \sim 100$ K), clearly highlighting the contrast. Remarkably, the extracted quasi-particle peak amplitudes scale logarithmically with temperature above $T^*$ and begin to saturate around $T^*$ (Fig. 2d), consistent with the typical behavior observed in Kondo systems[58,63].

The observation of Kondo behavior developing inside a magnetically ordered state formed by non-heavy electrons is interesting and distinct from the conventional picture, where the Kondo effect is typically suppressed by such magnetic ordering. This suppression occurs because magnetic ordering polarizes the conduction electron spins, thereby reducing the spin-flip scattering between the conduction electrons and *f*-local moments. To understand the unusual Kondo behavior in CeCo$_2$P$_2$, we calculate the temperature dependent *f-c* hybridization field, $\chi \sim \langle f^\dagger c \rangle$, in the AFM state. The calculation indeed shows that the hybridization starts to develop at $T^*$ and stabilizes at lower temperature (Fig. 2e), indicating the formation of a Kondo phase[56].

Although AFM ordering breaks time-reversal symmetry (*T*) in CeCo$_2$P$_2$, the combined inversion and time-reversal symmetry (*P•T*) is preserved[56,60]. The preservation of *P•T* symmetry, along with the unique lattice structure, facilitates the unusual Kondo phase observed in CeCo$_2$P$_2$. In each [P-Co$_2$-P] layer, ferromagnetic ordering results in the spin polarization of Co-3*d* conduction electrons. The *P•T* symmetry ensures that oppositely spin-polarized electrons from adjacent [P-Co$_2$-P] layers are degenerate. Consequently, the Kondo phase can develop through the formation of a novel type of Kondo singlet, involving cooperation between the two AFM layers. As illustrated in Fig. 2f, spin-down Ce-4*f* local moments are Kondo screened by spin-up conduction electrons from spin-up ferromagnetic order in one [P-Co$_2$-P] layers; while spin-up Ce-4*f* local moments are Kondo screened by spin-down conduction electrons from the spin-down ferromagnetic order in the adjacent [P-Co$_2$-P] layers.

In addition to the Kondo effect, our calculation reveals a band inversion occurs near $E_F$ between the Ce-4*f* and Co-3*d* bands, leading to the formation of a Dirac nodal-line structure protected by glide-mirror-*z* symmetry at $k_z = \pm 0.33$ Å$^{-1}$ in the bulk BZ (Fig. 3a). To investigate the nodal-line structure, we performed detailed photon energy-dependent ARPES experiments over a broad energy range of 300-800 eV, covering multiple BZs (Fig. 3a). The strong $k_z$ dependent characteristics can be seen in the $k_x$-$k_z$ spectral intensity map (Fig. 3a) and high symmetry dispersion along $k_z$ (Fig. 3b), demonstrating the bulk nature of the electronic structure.

The clear periodicity of the electronic structures along the $k_z$ direction (Fig. 3a and 3b), and the agreement between experiments and calculations (Fig. 3b and 3c), allows us to precisely identify the $k_z$ position of the nodal lines, which can be accessed using 380 eV photons (red lines in Fig. 3a). With the $k_z$ momentum identified, a series of band dispersions cutting through the nodal line (cuts 1 to 5, illustrated in Fig. 3a) are plotted in Fig. 3d. These dispersions exhibit crossings between the flat Ce-4*f* band and dispersive Co-3*d* band at $E_F$, forming nodal points that align well with the calculations shown in Fig. 3e. These nodal points collectively form the nodal line, as schematically illustrated in Fig. 3f. By tracking the band dispersions, the shape of the nodal line was determined experimentally in momentum space, as shown in Fig. 3g, again consistent with the calculations.

In addition to the unique bulk band structure that forms the Dirac nodal line, intriguing surface states are also observed in CeCo$_2$P$_2$. In Fig. 4, we compare the surface electronic structures to the bulk ones. The bulk band (Fig. 4a and 4b, blue color, indicated by dashed lines) forms an electron-type Fermi pocket by the Ce *f*-band, centered at the Γ point of the BZ. The photon energy dependence (and thus the $k_z$ momentum, the insets of Fig. 4a and 4b) pocket size (characterized by the double headed arrows) clearly demonstrates its bulk nature, in contrast to a recent study[64]. Unlike measurements with soft X-rays (Fig. 4b), where the surface states (red color in Fig. 4a) are strongly suppressed due to the enhanced bulk sensitivity, the surface states of $CeCo_2P_2$ are better visualized at lower photon energies, down to the VUV range (Fig. 4a, 4c and 4d). In this energy region, rich and complex surface states structures across the entire BZ can be clearly observed.

To understand these details, we performed *ab initio* calculations for both the bulk and surface bands [Fig. 4c (iii)] for comparison. Interestingly, only the shaded square pocket centered at the $\bar{\Gamma}$ point originates from the bulk state (blue), while all other pockets correspond to surface states (red), showing good agreement with the experiment [Fig. 4c (i, ii)]. In addition to the Fermi-surface topology, the experimental band dispersions of the surface states [Fig. 4d (i)] also match well with the calculations [Fig. 4d (ii)]. Notably, among different branches of surface states around the $\bar{M}$ point, one (marked by the red arrows in Fig. 4d) connects to the bulk nodal ring, forming drumhead-like surface state (Extended Data Fig. 7, 8 and 9). These surface states can modify the surface electronic environment and affect how conduction electrons screen the localized magnetic moments on the surface, potentially altering the surface Kondo screening in the presence disorder and perturbations, as well as the magnetic ordering at the surface.

The unique properties of $CeCo_2P_2$ make it an ideal platform for the study of rich physics phenomena, including: 1) a distinctive Kondo effect supported by $P•T$ symmetry, which robustly coexists with 2) an AFM state stabilized by the quantum geometry of a narrow Co $3d$-band, along with 3) the Dirac nodal line excitation formed by $f$-electron in the Kondo phase due to non-trivial band topology characterized by a glide-mirror-$z$ protection, and 4) rich surface states that interact with the unusual bulk electronic structures. Additionally, by lowering the magnetic crystalline symmetries, a variety of novel states can be realized, such as magnetic Weyl Kondo semimetal, axion insulator, and quantum anomalous Hall state (Extended Data Table III). Our results reveal the critical roles played by the wave functions of the electronic bands and the lattice symmetry in this strongly correlated electron system.

**Data availability**

All data are available in the main text.

**Acknowledgements**

We thank Fuchun Zhang, Jing Wang, Elena Hassinger for insightful discussion and Silke Paschen for the initial collaboration. We acknowledge Diamond Light Source beamline I05 (proposal nos. SI37110), I10 (proposal nos. MM32087), Bloch beamline of MAX IV (proposal nos. 20210939), BL5-2 of Stanford Synchrotron Radiation (Contract No. DE-Ac02-76SF00515), ADRESS beamline of the Swiss Light Source for access. This work is supported from the fundamental research funds for the central universities (Grant No. 10100-310400209541, 226-2024-00068, 226-2024-00200), National Key R&D Program of China (Grants No.2022YFA1402200), National Natural Science Foundation of China (Grant No. 12374454, 52331008, 12474044, 12374163, 12034017). B.A.B was supported by the Gordon and Betty Moore Foundation (Grant No. GBMF8685) towards the Princeton theory program, the Gordon and Betty Moore Foundation's EPiQS Initia- tive (Grant No. GBMF11070), Office of Naval Research (ONR Grant No. N00014-20-1-2303), Global Collaborative Network Grant at Princeton University, BSF Israel US foundation No. 2018226, NSF-MERSEC (Grant No. MERSEC DMR 2011750), Simons Collaboration on New Frontiers in Superconductivity and the Schmidt Foundation at the Princeton University.

**Author Contributions**

Y.L.C. and D.F.L. conceived this work. D.F.L., J.Y.L. performed the ARPES experiments with the assistance of C.C. and D. P. Y.F.X., H.Y.H., Y.J. and B.A.B. performed the theoretical calculations and analysis. T.P.Y., Y.Y.L., J.J.W., Y.B.C., H.H. synthesized and characterized the single crystals. J.Y.L., Y.H.Y., D.Prabhakaran, D.B., T.L.L., P.S., P.B., T.H., G.V.D.L. performed the XPS and XAS experiments and data analysis. Q.H.Z., F.Q.M., and L.G. performed the STEM measurements. B.T., C.P., M.H., D.H.L., N.B.M.S., V.N.S., A.L., C.Cacho, D.B., T.L.L., P.S., P.B. provided the beamline support. L.X.Y., Z.K.L. and H.Q.Y. contributed to the scientific discussions. D.F.L., B.A.B. and Y.L.C. wrote the paper with contributions from all authors.

**Competing interests**

The authors declare no competing interests.

**Figures:**

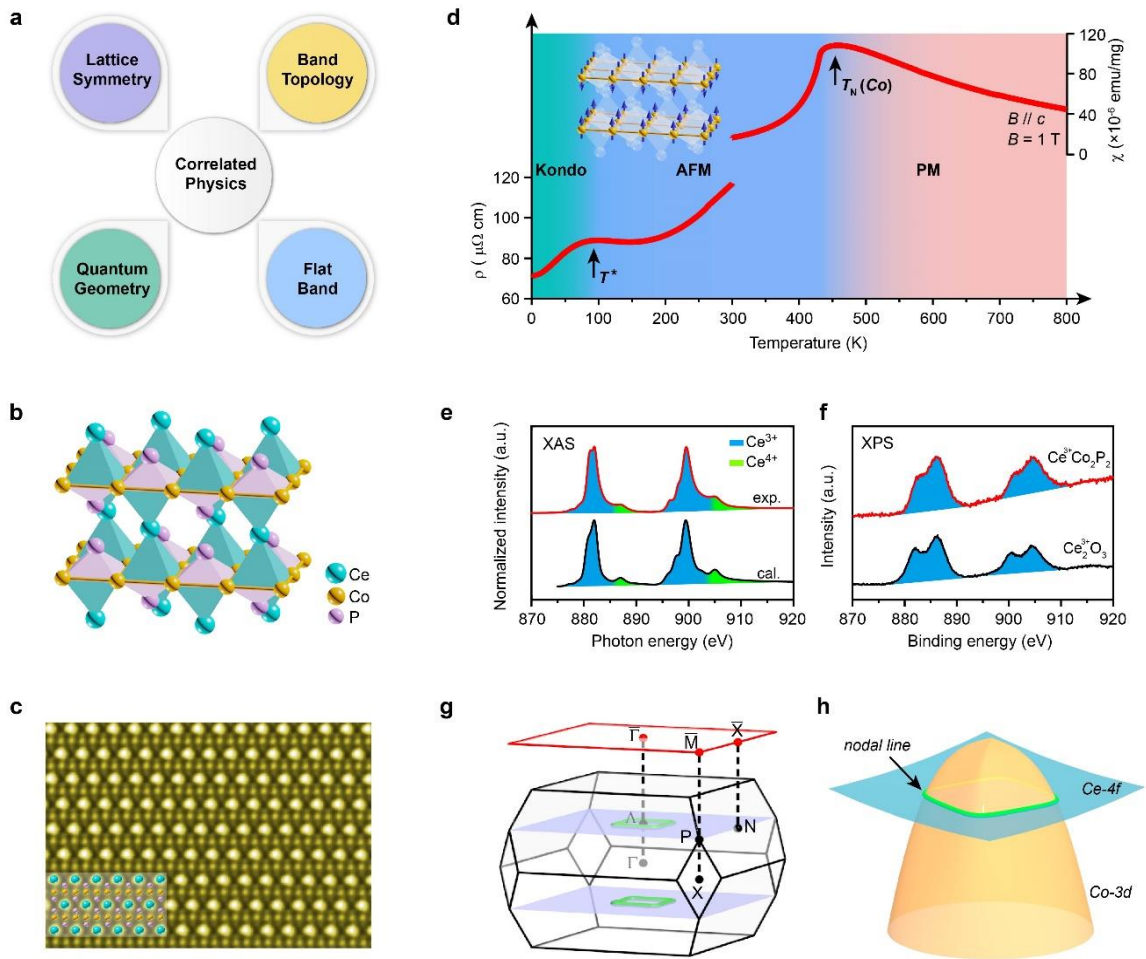

**Fig. 1 | Magnetic Dirac-Kondo semimetal phase in CeCo$_2$P$_2$. a,** The existence of rich exotic properties in CeCo$_2$P$_2$. **b,** Crystal structure of CeCo$_2$P$_2$. **c,** Atomic-resolution image of CeCo$_2$P$_2$, measured by high-resolution STEM. **d,** The phase diagram of CeCo$_2$P$_2$ as a function a temperature. The temperature dependent resistivity and magnetic susceptibility are overlaid. The antiferromagnetic transition ($T_N$) and the resistivity maximum ($T^*$) are indicated by arrows. The inset illustrates the magnetic structure of CeCo$_2$P$_2$, with the ordered magnetic moments formed on Co atoms. **e,** Comparison between the XAS spectrum of CeCo$_2$P$_2$ (red curves) and theoretical calculations. Ce$^{3+}$ and Ce$^{4+}$ peaks are indicated by blue and green shaded areas, respectively. **f,** XPS spectrum of CeCo$_2$P$_2$ showing four characteristic Ce$^{3+}$ peaks. The XPS spectrum of Ce$_2$O$_3$ is also shown for reference[59]. **g,** Bulk Brillouin zone (BZ, black lines) and the corresponding surface BZ (red lines) along the (001) surface of CeCo$_2$P$_2$. The position of nodal lines is indicated in green. **h,** Illustration of the Dirac nodal line formed by the inversion between Ce-4$f$ flat band and Co-3$d$ conduction band.

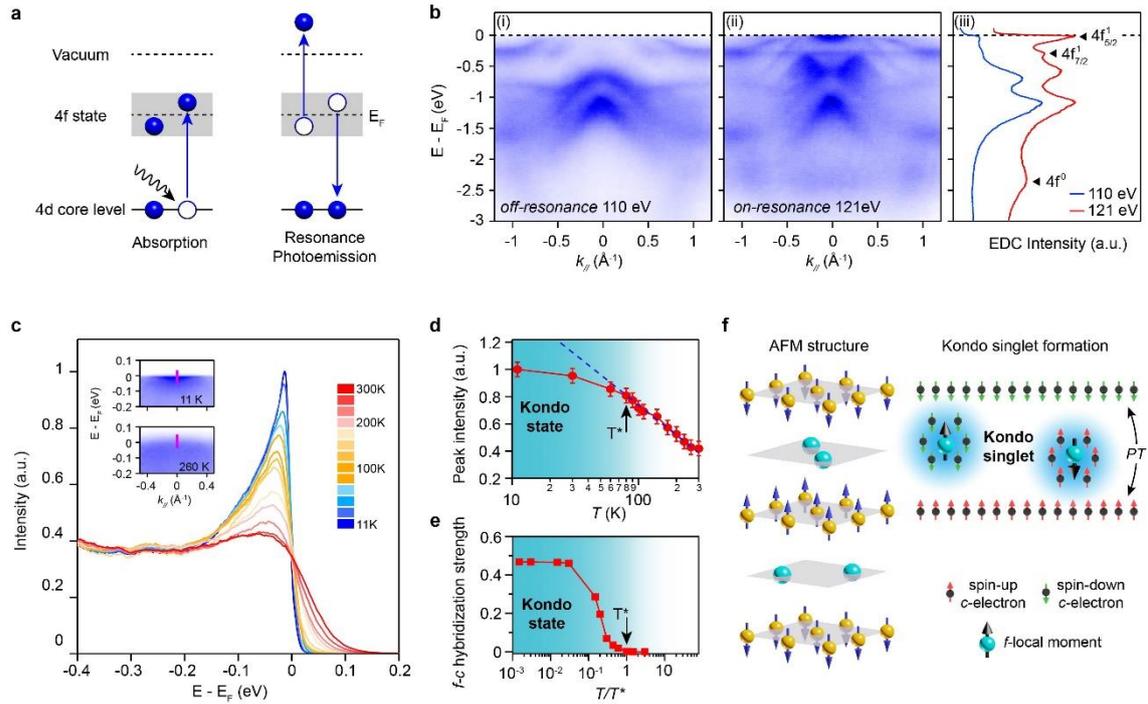

**Fig. 2 | Observation of the Kondo effect in CeCo$_2$P$_2$. a,** Illustration of the resonance photoemission process. **b,** Off-resonance (i) and on-resonance (ii) ARPES spectral data of CeCo$_2$P$_2$, and the corresponding energy distribution curves (EDCs) extracted at $k_{//}$ = 0. The Ce-4$f$ band positions are indicated by triangles. **c,** Temperature evolution of the Ce-$4f^1_{5/2}$ band at $E_F$. Insets show the Ce-$4f^1_{5/2}$ band at low and high temperatures, respectively. The EDCs extracted at $k_{//}$ = 0, as indicated by the pink lines in the inset. **d,** Temperature evolution of the Ce-$4f^1_{5/2}$ peak amplitudes on a logarithmic scale. The behavior follows -ln($T$) at high temperatures and starts to deviate at $T^*$. **e,** Calculated $f$-$c$ hybridization strength at different temperatures. It starts to develop at $T^*$, indicating the formation of Kondo state. **f,** Illustration of Kondo-singlet formation in CeCo$_2$P$_2$ inside the AFM state. The Ce-4$f$ local moments are Kondo screened by conduction electrons with opposite spin moments from different [P-Co$_2$-P] layers. Here, P atoms are not shown.

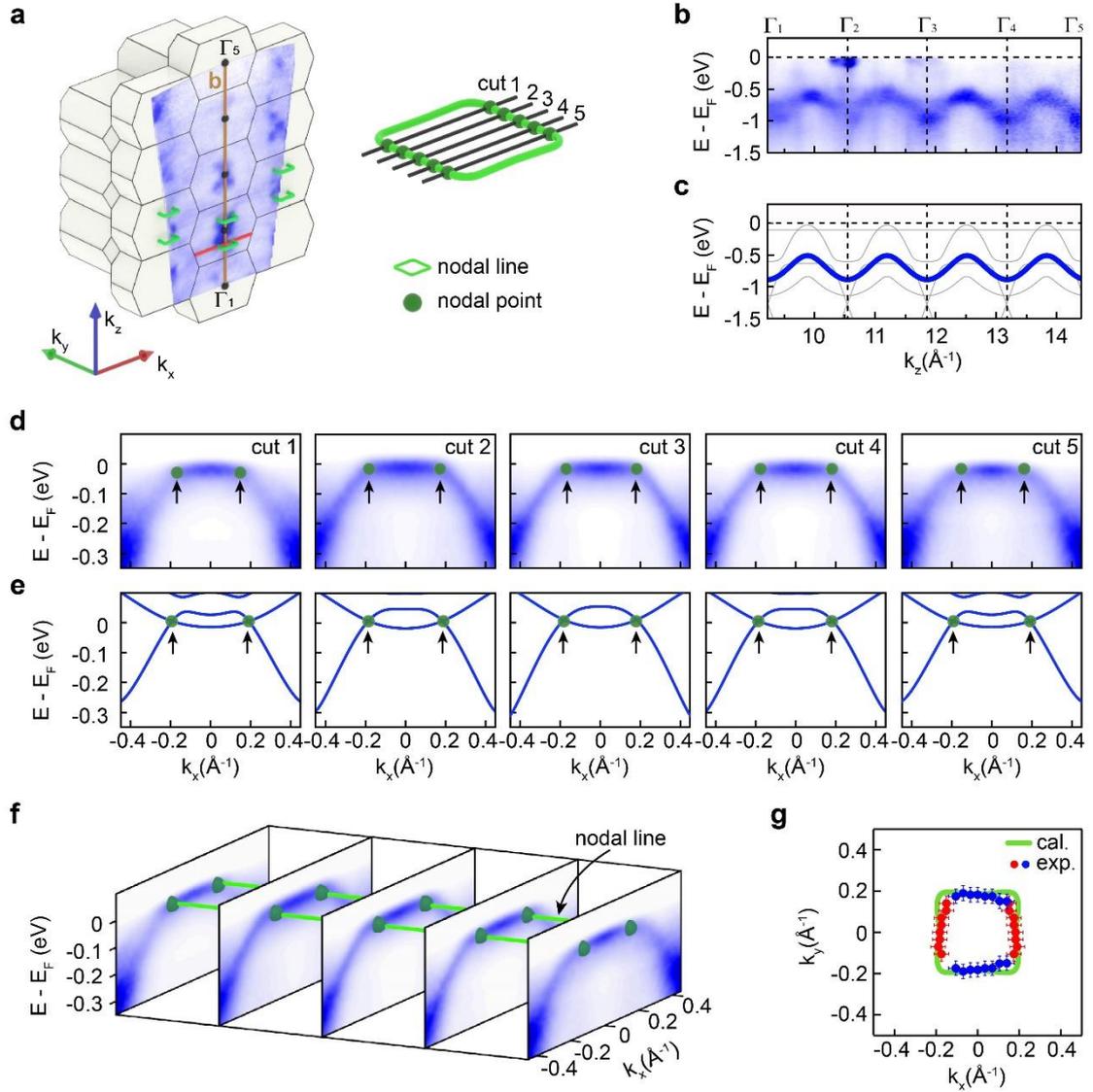

**Fig. 3 | Observation of the Dirac nodal line. a,** Schematic illustration of the multiple 3D BZs and the photoemission spectral map in the $k_x$-$k_z$ plane. The nodal lines (green) are selectively shown. **b,** Band dispersion along $k_z$ direction from $\Gamma_1$ to $\Gamma_5$. The momentum path is indicated by brown line in **a**. **c,** Calculated band dispersion in the paramagnetic phase along the $k_z$ direction, compared with the experimental results in **b**. The clear periodic dispersion in **b** between -1 eV and -0.5 eV corresponds to the blue highlighted band in **c**. **d,** Series of dispersions cutting through the nodal ring. The momentum path is illustrated in **a** from cut 1 to cut 5. **e,** Corresponding calculated dispersions in the AFM state to compare the experimental results in **d**. The nodal points are indicated by arrows. **f,** Stack plot of band dispersions at different $k_y$ values, showing the nodal line structure. **g,** The extracted positions of the nodal lines. The blue points are obtained based on the crystal symmetrization of the red points.

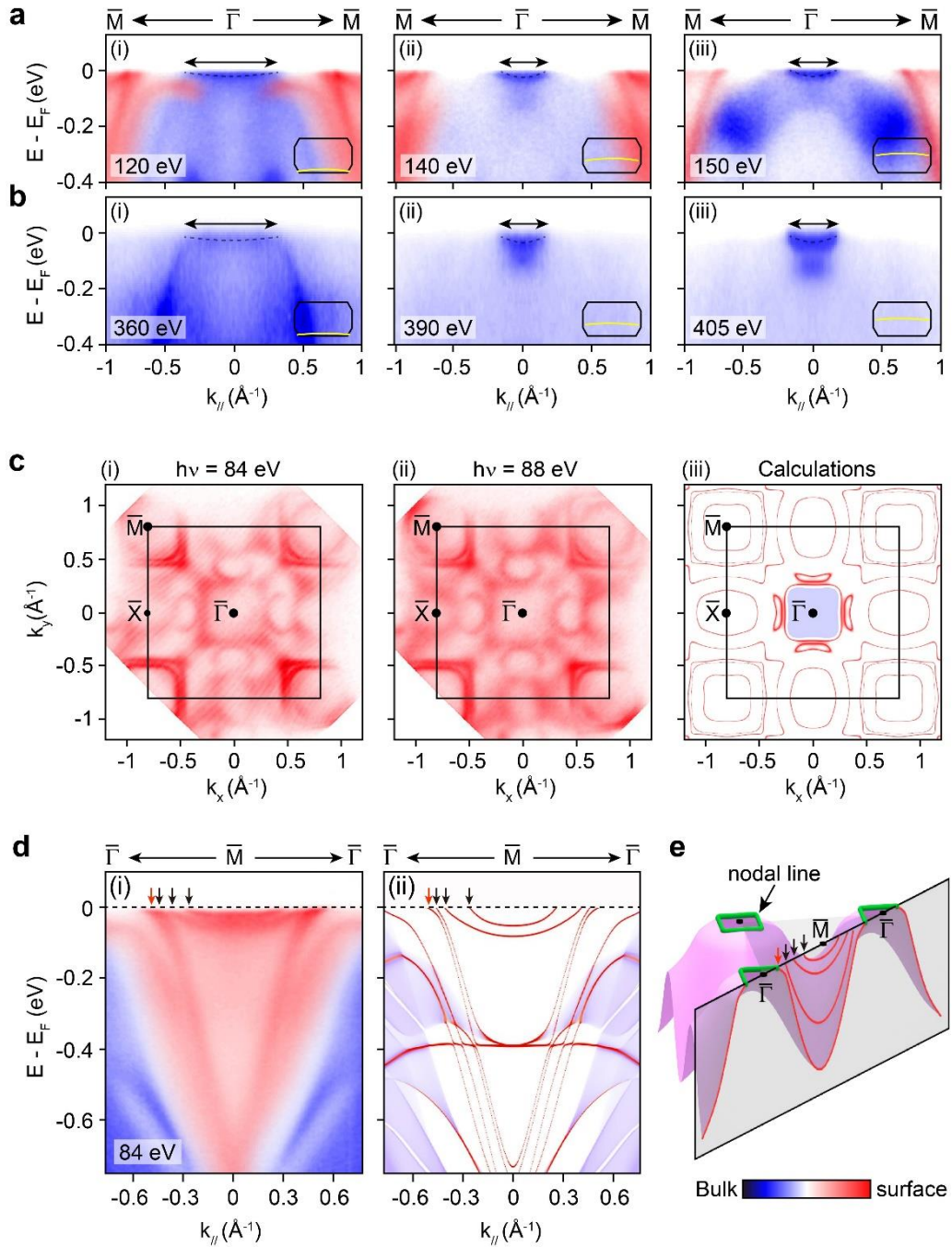

**Fig. 4 | Identification of the bulk Ce-*4f* band and surface states in CeCo$_2$P$_2$. a,** Photoemission spectral intensity along the $\bar{\varGamma}$-$\bar{M}$ direction taken at various photon energies. The red and blue spectral intensities represent the surface and bulk states, respectively. Insets show the corresponding $k_z$ momentum locations (yellow curves) in the BZ (black curves). The size of Ce-4*f* band is indicated by double headed arrows. **b,** Same as **a,** but taken using soft X-rays, where the surface states are suppressed. **c,** Comparison of the experimental (i, ii) and calculated (iii) FSs, showing good agreement. **d,** Comparison of the experimental dispersion (i) along the $\bar{\varGamma}$-$\bar{M}$ direction to the calculations (ii). **e,** Illustration of the surface states.